# Edge Intelligence-Driven LegalEdge Contracts for EV Charging Stations: A Fedrated Learning with Deep Q-Networks Approach

Rahim Rahmani[a]1 , Arman Chianeh[b]

[a]Department of Computer and Systems Sciences, Stockholm University, Sweden,(rahim@dsv.su.se)
[b]Department of Computer science, Tabriez azad University, (arman.research.dev@gmail.com)

**Abstract** We introduce LegalEdge, an edge intelligence-driven framework that integrates Federated Learning (FL) and Deep Q-Networks (DQN) to optimize electric vehicle (EV) charging infrastructure. LegalEdge contracts are novel smart contracts deployed on the blockchain to manage dynamic pricing and incentive mechanisms transparently and autonomously. By leveraging FL, multiple edge devices such as EV charging stations—collaboratively train DQN agents without sharing raw data, preserving user privacy while reducing communication costs. These edge-deployed agents learn optimal charging strategies in real time based on local conditions and global policy updates. LegalEdge ensures low-latency decisions, high contract integrity, and efficient energy allocation. Our experimental results demonstrate significant improvements in learning convergence, transaction speed, and operational transparency, establishing LegalEdge as a scalable, intelligent, and accountable solution for next-generation EV charging networks.

## 1. Introduction

The global push towards sustainable energy solutions has led to the rapid growth of electric vehicles (EVs). However, the widespread adoption of EVs presents new challenges in energy management and infrastructure optimization, particularly in ensuring efficient, secure, and transparent charging processes. Traditional energy management systems often lack flexibility, transparency, and verifiability, creating inefficiencies and potential vulnerabilities. Rapid growth when facilitated by policy support, mostly due to their multiple benefits in the areas of transport decarbonization, air pollution, and energy efficiency improvements. Falling battery costs and longer driving ranges, more vehicles available with fully or partly electric motors, and the deployment of charging infrastructure compounded this trend. Whenever they are connected to the electricity grid EVs are dependent on a charging station at home at work or at a public charging station. In the future, different options for load management should become possible concerning charging with normal and medium power in both residential and commercial areas. Triggered by new technologies, policy, regulation and customer expectations are all contributing to a shifting paradigm such that EVs were charged in an uncontrolled way [1] they could increase the peak on the grid since charging trends could match existing load peaks and thus contribute to overloading and the need for upgrades at the distribution and transmission levels. It is getting harder and more costly to predict and balance demand and supply while lacking visibility into millions of new consumer devices and distributed energy resources popping up at the grid edge [1] in their work classified EV charging into three types namely, uncontrolled charging, delayed charging, and smart charging. Uncontrolled charging is a type of charging where an EV user plugs their vehicle, and it starts charging until it is fully charged, or the owner disconnects it later. Delayed charging is the type of charging where utility companies use the tariff to motivate EV owners to shift their EV load from peak load hours to off peak hours [1]. Smart charging is a type of charging in which the EV charging point is controlled by an algorithm. At present EV charging often takes place during existing electricity system peak times (4-9 PM). The simultaneous charging of many EVs at these times increases the peak demand on the grid, potentially causing overloading of different grid components and congestion and may thus require additional infrastructure investment at the distribution level [2], [3], [4], [5], [6], [7]. The increase in electricity demand may also result in additional generation capacity needs which would require additional investments to ensure adequate electricity supplies. These challenges of integrating EVs into electric grids can be addressed through managed EV charging also known as smart charging. Smart charging allows EVs to be more smoothly integrated to the grid with reductions in load timing problems. Smart charging implementations include: I) Unidirectional power flow management during EV charging (V1G) enables EVs to shift the time of day of charging to better fit with grid conditions and II) Bidirectional power flow known as vehicle-to-grid (V2G) enables EVs to charge from the grid or discharge energy stored in their batteries to the grid to go beyond simply just being able to modify the times that EVs charge but to effectively use EVs as stationary electricity storage devices. Blockchain technology has emerged as a promising solution to address these challenges. Smart contracts provide automated, tamper-proof mechanisms for executing predefined conditions, while Ricardian contracts bridge the gap between legal agreements and digital automation. Integrating these technologies within the EV charging ecosystem offers the potential for enhanced efficiency and accountability. In [8] proposed a power transaction architecture based on a multi-agent alliance which uses blockchain as the contract settlement system. In [9] a new hybrid blockchain storage mode was designed and in [11] proposed a blockchain-based edge-as-a-service framework for secure energy trading in software-defined networking (SDN) enabled vehicles to grid environments. The power transactions in the energy Internet proposed by [8][9][11][12] do not involve the power transaction of EVs. The blockchain-based EV charging system will allow the billing and payment functions by smart contract deployed on the blockchain platform. Edge intelligence, which leverages distributed computing resources at the network's edge, further enhances the system's capabilities by enabling real-time data processing and decision-making. When combined with reinforcement learning, edge intelligence can optimize charging operations dynamically, learning from past interactions to improve future performance. A smart contract can be used to reward EV users for their flexibility and offers to incentivize them to use smart charging. Distributed Edge Intelligence (DEI) ability to provide context to the collected raw-data, i.e. contextualization of the raw-data; minimizing dependency for executing tasks due to closeness to data sources and the DEI will improve performance of tasks through experiences, i.e. learning; predicting an outcome in the event of uncertainties; efficient routing of the data; self-organization of the things. DEI is designed to process large amounts of data in a distributed way and make up for the lack of cloud computing capabilities in different domains such as smart city, eXten Reality(XR) and distributed charging station to deploy smart contracts. In order to realize the dynamic and adaptive [13] management and scheduling of the Ede the reasoning and learning ability of AI is needed. DEI can be characterized in edge-to-edge collaboration, Edge autonomy and resource elasticity. To address the issues of charging stations in urban and remote areas due to sparse charging stations and the passive scheduling of those stations, this paper proposes a coherent blockchain based algorithm for distributed charging stations by Intelligence-enabled Edge computing to push the distributed data processing in the smart contract necessitates by that more powerful methods, i.e., AI/Machine Learning (ML) technologies for extracting insight data and knowledge acquisition that lead to better decisions and strategic business moves.

This paper proposes a novel framework that integrates Ricardian contracts, smart contracts, and edge intelligence powered by fedreated learning to create an advanced EV charging station management system. The contributions of this work include:

    1.A blockchain-based architecture for secure and transparent policy execution using Ricardian and smart contracts.

    2.The incorporation of edge intelligence for real-time decision-making and operational efficiency.

    3.The application of federated learning to optimize charging processes and resource allocation.

By addressing the limitations of traditional systems and leveraging cutting-edge technologies, this research aims to provide a scalable and new solution and introduces LegalEdge Contract a scalable, secure, and intelligent solution designed for the next generation of EV charging stations. LegalEdge Contract as shown in figure 1 is an innovative and forward-thinking contractual framework that seamlessly merges the structured, legally enforceable nature of Ricardian Contracts (RC) with the automated, decentralized execution of Smart Contracts (SC). This hybrid approach ensures that agreements are both human-readable and machine-executable, offering a scalable, secure, and intelligent solution for modern digital transactions.

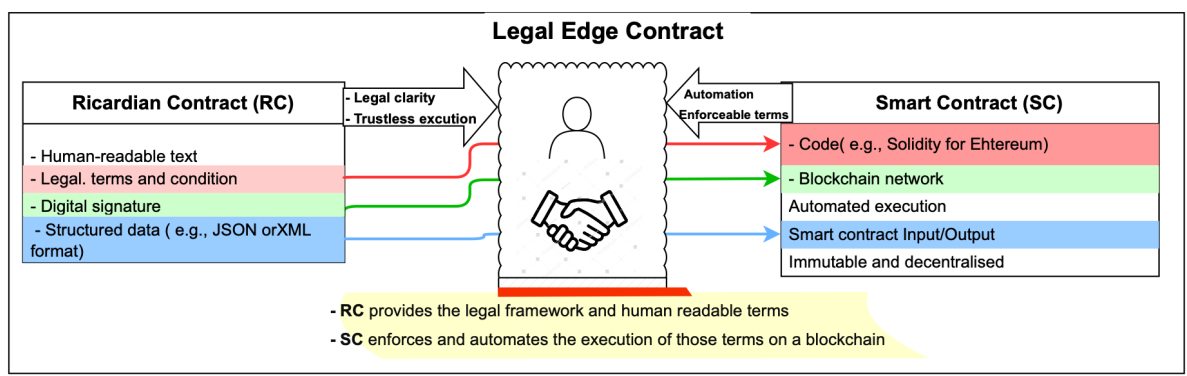

    Figure 1. Legal Edge Contracts

## 2. Related Work

The aim of this section is to review recent works in EVs charging system with application of DEI in energy systems. The emphasis is on load management and flexible trading on a relatively fast time scale while the issue of load planning and scheduling has been considering as part of load management and flexible trading.

In the [Okwuibe] describe multiple actors in the eclectic vehicle charging market. In figure 2 the traditional market, its actors, and its communication flows are shown. EV users often communicate via one or multiple third parties with the CPO [13]. These third parties include an E-Mobility service provider that bridges the gap between EV and CPO to provide information about existing CS and an E-Mobility clearing house that handles payments between EV and CPO [13]. The CPO needs to engage with an electric supplier to buy electricity from the energy market. Electricity needs to pass a distributed system operator (DSO), more commonly known as a utility company [13] to get electricity. The DSO operates a medium or low voltage (MV/LV) grid, through which energy from the energy markets distributed energy resources (DER) can pass through to the CS and into the EV.

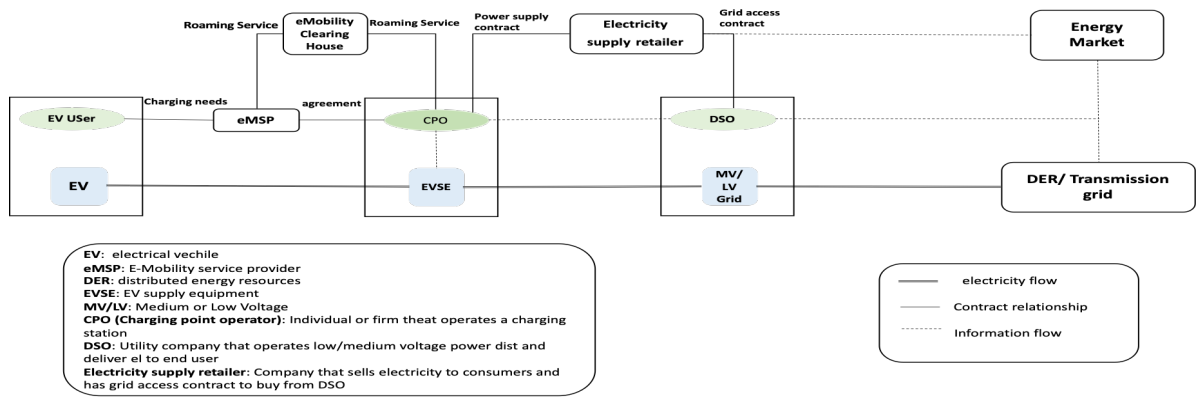

Fig. 2 EV Charging Market architecture [13]

This architecture faces some challenges. The current solution is mainly centralized and often based on a request-reply architecture [14][15]. This poses problems with scaling, as it can be difficult to scale such solutions well [16]. It also introduces a single point of failure into the system, as third parties and their solutions must work for all charging with a CPO to function [14]. There is also a power disadvantage, as the CPO and third parties own the data generated by the EV and EV users [14][15]. This power disadvantage gives the centralized authorities much power over users' data and how they are used [14][15]. If a vulnerability is exploited, this private information could leak, invading users' privacy [14][15] describes that habits and patterns, such as working hours, workplace, and frequently visited areas, can be derived from the data stored by CPO and third parties. This information could be leaked and used for targeted advertisement [14].

There are also reports of trust issues, as all the components in the charging process must function for charging to be correctly executed, [14] describes that correct charging logic is often not correctly implemented, leading to incorrect operations and prices. Blockchain technologies can improve the current situation by removing the need for a third party [17] as shown in figure 3 and facilitating communication between the parties using smart contracts on a blockchain [13], [18],[19].

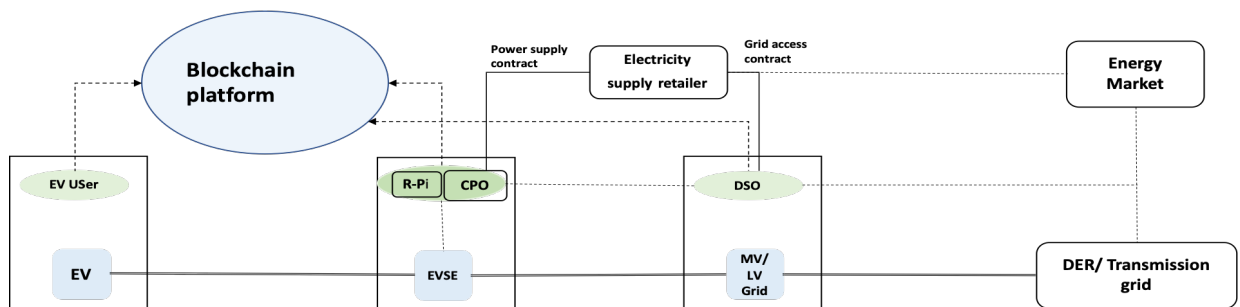

Fig. 3. EV charging market architecture with blockchain

Smart contracts are a self-executing program stored on the blockchain [20]. These programs are immutable and cryptographically verified to ensure their trustworthiness [20]. In general, smart contracts are executed when one of its defined functions is called upon inside of a transaction [20]. Executing a smart contract function is similar to executing any other transaction on the blockchain: peer- to-peer without any centralized party [20]. Ethereum was among the first blockchain systems to introduce a smart contract that was Turing complete [21]. This means that Ethereum smart contracts can run any algorithm [22]. Blockchain makes these programs tamper-proof and self-executing [23], making them ideal for trusted communication between untrusted parties [20]. Untrusted parties can safely communicate via a smart contract without needing a third party [20]. Eliminating third parties benefits all involved parties, as it removes the cost and a centralized authority associated with third parties [20]. Because the smart contracts are self-executing, no single party owns the execution and can thus trust that it will operate according to its programming [20]. A big challenge for blockchains and their smart contracts is that they are blind to the real world [24]. The only way for data from the real world to get injected into a smart contract is if an actor from the real world sends such data [24]. The solution for smart contracts to get information about the real world is using an Oracle [24]. An Oracle can be a centralized and trusted third party that injects data into the smart contract from the real world [24]. Relying on Oracles is thus often seen as a problem, as it reintroduces the need for centralized trusted third parties for the specific function that needs real-world data [24]. As discussed above the blockchain charging systems research is scattered into different focuses and scenarios. To grasp the current state of research and how our research builds upon them, the different areas where blockchain has been used to improve traditional charging systems need to be experimented by adding DEI and a distributed learning procedure that learns the optimal values for all weight and bias or the hidden patterns. To depict an active distribution electricity network system aligned with the concept of Active Distribution Network (ADN) proposed by the International Council on Large Electric Systems (CIGRE) [25], we utilize distributed edge computing and intelligence. This approach enables real-time decision-making, enhances system resilience by mitigating the risks of single-point failures, and addresses concerns related to privacy leakage. ADN is characterized by features such as high penetration of distributed renewables sources and energy storage systems prevalence of EV, participation of flexible loads and formation of microgrid /virtual power plants [26]. There is need to design an effective EV trading mechanism in ADN to achieving of "shaving peaks and eliminating valleys" [27]. In [28], blockchain technology is utilized to develop a secure peer-to-peer (P2P) electricity trading system. The study focuses on applying this method to electric vehicle (EV) charging systems through a decentralized trading model. However, it lacks the integration of distributed edge intelligence, which is essential for enabling real-time decision-making and enhancing system responsiveness. In [29], distributed edge intelligence is employed to develop a hybrid charging scheduling method within the Internet of Vehicles (IoV). The study focuses on determining the optimal charging mode for EVs. However, it does not address the EV trading mechanism, which significantly impacts charging costs. Given this 2–3 orders of magnitude gap, this research aims to address the following question: How can the EV charging system be enhanced through distributed edge computing by integrating blockchain and AI technologies? The focus is on leveraging the powerful reasoning and learning capabilities of AI within edge intelligence to optimize the system's efficiency, security, and decision-making processes.

## 3. Problem Formulation Optimizing EV Charging Processes Using Legal Edge Contract

The increasing adoption of EVs presents significant challenges in ensuring efficient, secure, and transparent charging processes. Traditional EV charging systems often suffer from issues such as lack of trust between stakeholders, fragmented payment systems, inefficient energy distribution, and regulatory complexities. Moreover, the absence of a unified framework that integrates legal agreements with automated execution creates barriers to scalability and interoperability.

To address these challenges, this research introduces LegalEdge Contract, a novel framework that merges Ricardian and smart contracts to optimize EV charging operations. LegalEdge Contract ensures legally binding agreements between stakeholders while leveraging blockchain-based automation to streamline payments, enforce terms, and enhance security. By integrating structured legal frameworks with decentralized execution, this approach enables:

- Trustless and transparent transactions between EV owners, charging stations, and grid operators.
- Automated energy management to optimize grid load balancing and dynamic pricing.
- Regulatory compliance and legal enforceability, ensuring adherence to policies and standards.
- Seamless payment processing through tokenized transactions and digital wallets.

The problem formulation thus revolves around designing and implementing LegalEdge Contract to create a scalable, secure, and intelligent EV charging ecosystem that enhances efficiency, reduces operational risks, and fosters widespread adoption.

## 4. System Architecture

In this section we introduce definitions sysetem architecture, data flow and processing, distributed algorithm and AI models and simulation and or experimental design.

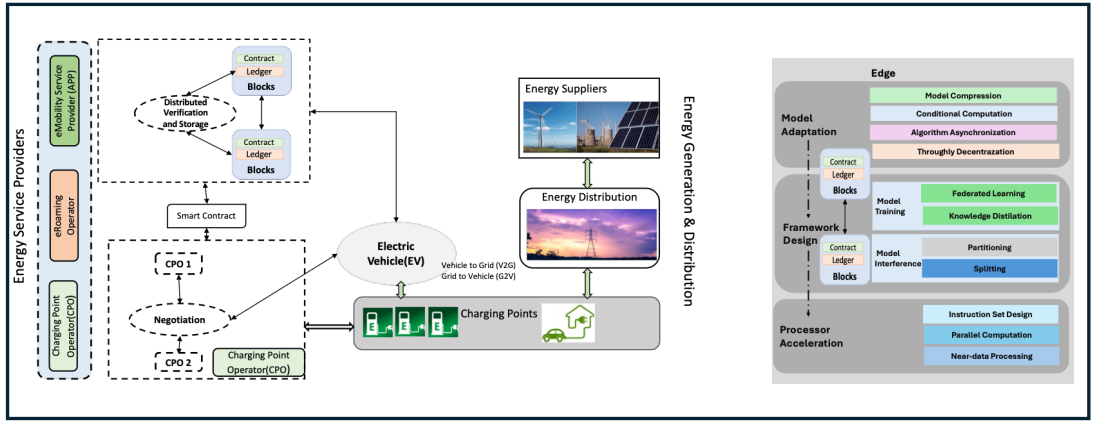

Fig. 4 (a) System Architecture Edge-based EV Charging with blockchain (b) Edge Intelligence extended of [12]

- **System Architecture:** As shown in figure 4 (a) and (b) the system architecture to enable Fedrated Learning (FL) and blockchain at the edge. The proposed solution consists of multiple features in the distributed edge-based architecture for EV charging. Include key components like edge nodes, charging stations, data acquisition systems, and communication protocols. The edge nodes host the components for identify EV to interact with, Identification will allow edge node to authorise actions the EV wants to achieve, such as charging. The edge node identifies CPOs that interact with it. Identification will allow the edge node to authorise actions the CPO wants to achieve, such as registration CS. The edge node must identify CSs that interact with it. Identification will allow the edge node to authorise actions the CS wants to achieve, such as discharging. The edge node should facilitate agreements: I) between EV and CPO, II) with variable length, III) with energy source selection and IV) with maximum charging rates.
- **Data Flow and Processing:** A smart contract is a program residing on a blockchain that can independently execute its functions [31]. An overview of the smart contract and its environment is illustrated in Figure 5.

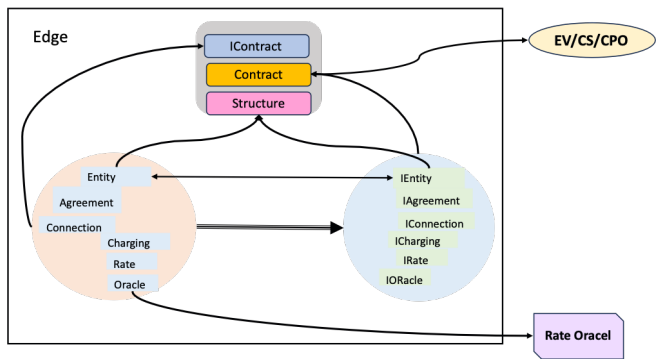

Fig. 5 Data flow and processing

The framework primarily interacts with two key components:
**I)Entities:** These include the EV, CS, and CPO, which directly communicate with the framework.
**II)Rate Oracle**: This component provides the current charging rates to be applied.
This simplified overview highlights the artefact and its interactions with other components.

**4.1 LegalEdge Contract:** A Hybrid Framework for Intelligent, Secure, and Scalable Agreements - Edge devices for real-time data processing

*Core Principles of LegalEdge Contracts:*

A. **Legal Clarity with Embedded Legalese**
LegalEdge Contracts incorporate natural-language legal terms that define the agreement's obligations,

rights, and conditions. This ensures that all parties including regulators, businesses, and individuals can understand the contract's meaning without requiring extensive technical expertise.

B. **Smart Automation and Execution**
While Ricardian Contracts provide the legal backbone, LegalEdge Contracts integrate Smart Contracts to enforce and automate key contractual clauses. This ensures that transactions are trustless, self-executing, and immutable, reducing the need for intermediaries.

C. **Blockchain-Enabled Trust and Security**
LegalEdge Contracts leverage blockchain technology to enhance transparency, security, and tamper resistance. Each contract is cryptographically signed and stored on a distributed ledger, ensuring its authenticity and resilience against fraud.

D. **Dynamic and Adaptive Compliance**
One of the key innovations of LegalEdge Contracts is their ability to adapt to regulatory changes. Using oracles and AI-driven legal analysis, they can update compliance requirements dynamically while maintaining legal integrity.

E. **Interoperability and Scalable Deployment**
LegalEdge Contracts are designed to be platform-agnostic, supporting various blockchain networks and enterprise systems. This makes them ideal for applications such as:
  - Finance & Tokenization: Automating loan agreements, digital asset issuance, and regulatory reporting.
  - Supply Chain & Logistics: Ensuring contract enforcement across distributed suppliers.
  - Healthcare & Data Privacy: Enabling privacy-preserving data sharing under regulatory frameworks like GDPR.
  - Energy & IoT Integration: Facilitating smart grid transactions, billing, and compliance.

F. **Human-AI Collaboration in Contract Execution**
LegalEdge Contracts integrate AI-driven legal reasoning to enhance contract negotiations, risk assessment, and dispute resolution. This ensures that contractual obligations remain both technically enforceable and legally sound over time.

## 4.2 Using Discrete Finite Automata for LegalEdge Contracts

LegalEdge Contracts can be modeled using Discrete Finite Automata (DFA) to ensure deterministic, rule-based execution of contract states. A DFA provides a structured way to represent contractual workflows, transitions, and enforcement mechanisms, ensuring that each contract execution adheres to predefined rules while integrating automation and compliance. Formaly, using [40] notation , the DFA specifies a computation as a 5-tuple, $(Q, \Sigma, \delta, q_0, F)$ where:

  i. A finite set of states, denoted $Q$
  ii. A finite set of input symbols (Events) called the alphabet ($\Sigma$)
  iii. A transition function ( $\delta: Q \times \Sigma \rightarrow Q$)
  iv. A start state ($q_0 \in Q$)
  v. A set of accept (end) states ( $F \subseteq Q$)

*4.3 DFA Representation of LegalEdge Contracts*

  i. **States (Q):**
       - *Drafted:* The contract is created with embedded legalese and smart contract terms.
       - *Signed:* All parties approve the contract, cryptographically signing it on the blockchain.
       - *Active:* The contract is in effect, and transactions can be executed based on predefined conditions.
       - *Triggered:* A contractual clause is activated due to an event (e.g., payment due, compliance check).
       - *Updated:* Adaptive compliance mechanisms modify terms in response to regulatory changes.
       - *Disputed:* A disagreement arises, triggering AI-driven legal reasoning for resolution.
       - *Completed:* All obligations are fulfilled, and the contract is successfully closed.
       - *Terminated:* The contract is canceled due to violation, expiration, or external intervention.
  ii. **Alphabet (Σ):** Events that trigger state transitions
       - *Signing request received* (transitioning from Drafted → Signed)
       - *Contract execution triggered* (Signed → Active)
       - *Transaction condition met* (Active → Triggered)
       - *Regulatory update detected* (Active → Updated)
       - *Dispute filed* (Active → Disputed)
       - *Resolution reached* (Disputed → Active or Terminated)

- *All obligations met* (Active → Completed)
- *Violation detected* (Active → Terminated)

iii. **Transition Function (δ):**
Defines how the contract moves between states based on events. For example:
- δ(Drafted, Signing request received) → Signed
- δ(Signed, Contract execution triggered) → Active
- δ(Active, Transaction condition met) → Triggered
- δ(Active, Regulatory update detected) → Updated
- δ(Active, Dispute filed) → Disputed
- δ(Disputed, Resolution reached) → Active
- δ(Active, All obligations met) → Completed
- δ(Active, Violation detected) → Terminated

iv. **Initial State (q0):**
- q0 = Drafted (Every contract starts in the Drafted state)

v. **Final States (F):**
- Completed (Successful contract execution)
- Terminated (Contract failure or early termination)

*4.4 Advantages of DFA for LegalEdge Contracts*

- **Formal Verification:** DFA ensures that every contract follows a well-defined execution path, preventing unexpected behaviors.
- **Automation & Smart Execution:** Automates contract enforcement without manual intervention.
- **Regulatory Adaptability:** The contract can transition to an *Updated* state dynamically in response to new laws.
- **Dispute Resolution Workflow:** Clearly defined paths help integrate AI-driven legal analysis for resolving conflicts.

This DFA-based approach ensures that LegalEdge Contracts are **secure, transparent, and deterministic**, aligning with the goal of **scalable, intelligent, and compliant contract execution** across industries.

## 5. Methodology

In this section we introduce definitions sysetm architecture, data flow and processing, distributed algorithm and AI models and simulation and experimental design.

**5.1 Implementation on the Ethereum Blockchain**

To test and evaluate the smart contract, a blockchain platform needed to be selected. Ethereum was chosen due to its widespread adoption and its support for Turing-complete smart contracts [30]. Ethereum's Ethereum Virtual Machine (EVM) is the execution environment for smart contracts [31],[32],[33], [34], and Solidity, one of the most popular EVM programming languages [33], was deemed appropriate for this implementation. The development process included implementing the following components in a simulated environment: I)Entities (EV, CS, and CPO) and II)Rate Oracle**:** These components were implemented using JavaScript and the widely used Web3JS library [35].

**5.2 Modular Smart Contract Design**

The Ethereum network imposes a size limit for each deployed smart contract (24576 bytes) as specified in Ethereum Improvement Proposal (EIP) 170 [36],[37]. Due to this limitation, the entire model could not be implemented within a single contract. Instead, it was split into multiple smaller contracts, each corresponding to specific components of the model. As shown in Figure 5, the structure consists of interconnected minor contracts. One of these contracts acts as an entry point, facilitating communication among the components and ensuring the execution of the correct functionality. This modular approach enhances coherence and scalability in the smart contract implementation.

- **Distributed Algorithms and AI Models**: The algorithms and AI models used in edge intelligence for decision-making, predictive maintenance, load balancing, demand forecasting and optimization. Figure 6 shows a user case of 3 edge nodes which shows the process of the distributed algorithms with FL and blockchain. To introduce the process, we assume that there are *N* edge nodes and one edge coordinators which aggregate models.

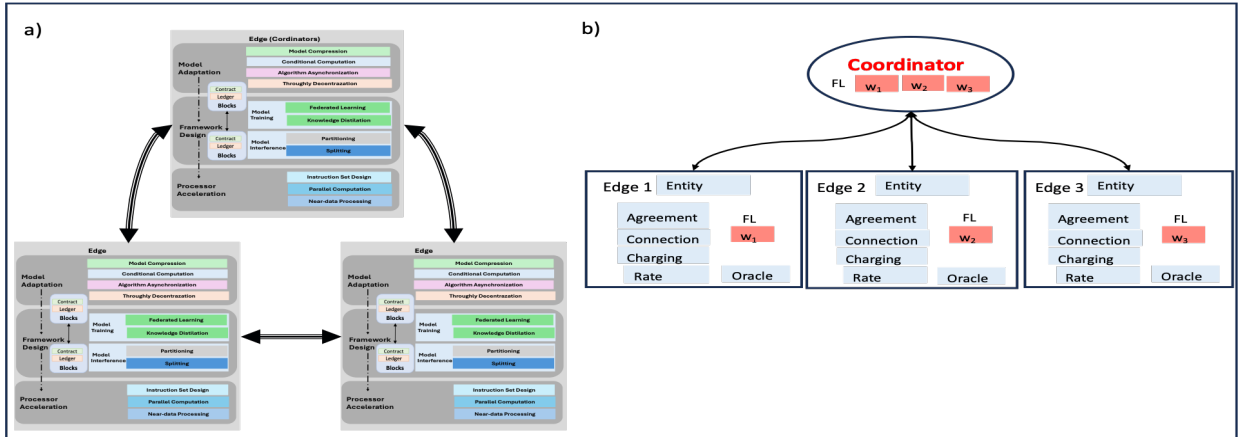

Fig.6 a) Distributed Edge intelligence with FL and blockchain b) FL data flow processing

With use of decentralized blockchain and integration FL in the process of charging process in distributed edge networks allows to eliminate the need for a central server in smart contract. Assume that *N* distributed edges and $d_n$ size of data captured on edge device *E*, and $E_N = |d_n|$ the number of samples on edge *E* in each edge train local data model $f_i$ with a function defined as $f_i = argmin\ F_i(w_i)$ där *w* is the aggregated central model and $F_i$ is the loss function. The coordinator aggregate (which is rate oracle as explained in the previous section) global model *w* with global objective function:

$$F(\omega) = \sum_{n=1}^{N} \frac{E_n}{M} f_n(\omega) \text{ where, } M = \sum_{n=1}^{N} |d_n| \qquad (1)$$

**5.3 Simulation and Experimental Design:** As outlined in Section 2, traditional research designs often rely on an additional entity to inform the smart contract about the charging process. The most common approach involves using an Oracle at each charging station (CS) to function as a "smart meter." This Oracle records when charging starts, when it ends, and the amount of electricity consumed—critical data for enabling secure and automated payments. However, relying on a third party like an Oracle introduces trust and reliability concerns. Electric vehicles (EVs) might underreport usage to reduce costs, while CS operators (CPOs) might overreport to inflate revenues. Furthermore, if charging begins before a deposit is made, payment enforcement becomes problematic, as the EV could refuse to pay after charging. To address these issues, the smart contract must act as an intermediary, holding the EV's deposit and disbursing the appropriate payment to the CPO after verifying the charging details. While the use of an Oracle or "smart meter" might appear practical, it introduces additional failure points and complexity to the system. A more robust approach involves calculating the necessary parameters before charging begins and drafting a predefined charging schedule for the EV and CS. This schedule addresses both issues by determining the required deposit amount for the EV and the precise quantity of electricity to be delivered. In order to venturing into the theoretical backbone of model efficiency and deployability next section explain Quantization in Federated Learning.

**Quantization in Federatedm Learning**

Quantization involves reducing numerical precision (e.g., float32 → int8) to:
- Lower memory footprint
- Accelerate inference (especially on edge/IoT hardware)
- Minimize communication overhead in FL

Quantization-Aware Training (QAT) simulates these constraints during training so the final model is robust to quantized inference. Subsection 6.1.3 mathematically demonstrates how LegalEdge's FL framework with DQN agents inherently enables QAT, particularly optimizing edge device inference.

## 6. Experimental Setup

The following section describes the experimental environment, including hardware and software configurations, as well as the simulation parameters for Federated Learning (FL). Figure 7 shows an illustration of the experimental systems setup.

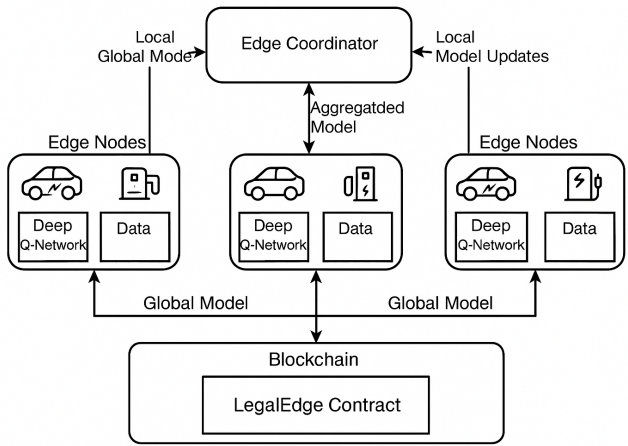

Figure 7. illustration of the experimental systems setup.

### 6.1 Experimental Environment Overview

#### 6.1.1 Objective
To simulate an EV charging ecosystem where smart contracts (LegalEdge Contracts) coordinate trustless energy transactions between EVs and Charging Stations (CSs) using FL and Deep Q-Networks for intelligent scheduling and load optimization. The architectural diagram showing in figure 8 how flower-based federated Learning coordinates DQN agents on edge nodes (EVs/Charging Stations), which interact with a blockchain hosting LegalEdge Smart Contracts.

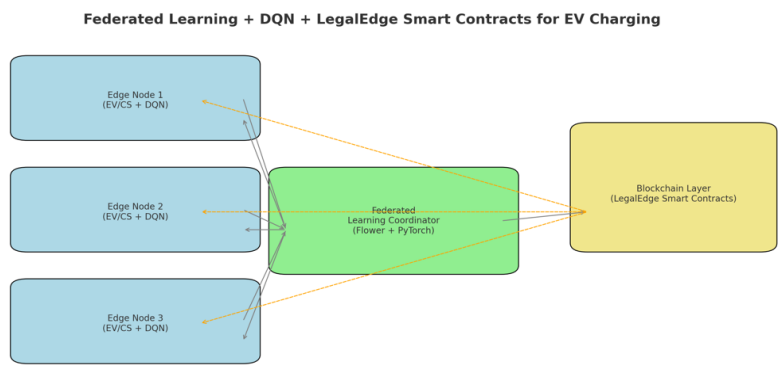

Figure 8. Flower-based federated Learning coordinates DQN agents on edge nodes

#### 6.1.2 Architecture

- **N Edge Nodes**: EVs, CSs, or local controllers, each with embedded AI.
- **1 Edge Coordinator**: Aggregates models from edge nodes in each FL round.
- **Blockchain Layer**: Hosts LegalEdge smart contracts ensuring trust and enforceability.
- **Federated Learning Layer**: Enables decentralized training of DQN agents across edge nodes.
- **RL Layer**: Each node uses a DQN to optimize its local charging policy (e.g., price, time, energy).
- **Simulation Agent Roles**:
  - EV Agent: Optimizes charging based on price, availability, state-of-charge (SoC), and contract terms.
  - CS Agent: Optimizes scheduling and pricing strategies.

o   Contract Agent: Executes LegalEdge contract logic (escrow, penalties, enforcement).

**6.1.3 LegalEdge FL with DQN and Natural QAT Support**

LegalEdge enables the QAT in both theory and design.

**1. DQN's Policy Representation is Value-Based**

A DQN learns a Q-function:

$$Q(s,a;\theta) \approx E[R_t | s_t=s, a_t=a]$$

Where:

- $\theta$ are neural network weights
- The final layer outputs a vector $Q \in R^A$ (one value per action)

This value vector is used to select actions:

$$a^* = arg^a max Q(s,a)$$

**Key Insight**: This argmax operation is robust to low-precision errors in Q-values and making it naturally tolerant to quantization.

**2. Federated Aggregation Works on Compressed Weights**

In LegalEdge-FL:

- Each client trains a local DQN and sends model updates (weights or deltas) to the server.
- These can be quantized:
  $\hat{\theta}_i = Q_{int8}(\theta_i)$

  where $Q$ is a quantization operator.

  Using Federated Averaging:

  $$\hat{\theta}_i = \frac{1}{N} \sum_{i=1}^{N} \hat{\theta}_I$$

This aggregation still converges under stochastic quantization:

$E[\hat{\theta}] \approx \theta^*$

**Theoretical Justification**: Under mild assumptions, quantized *SGD/FedAvg* preserves convergence in expectation:

$$\lim_{t \to \infty} E[L(\theta_t)] \to L^*$$

**3. Quantization-Aware Modules During Training**

LegalEdge uses PyTorch-based DQNs. PyTorch provides:

torch.quantization.prepare_qat()
Which:

- Inserts quantization layers into training
- Simulates 8-bit arithmetic using floating-point emulation
- Adjusts weight scales and zero-points

This is **natively compatible** with LegalEdge's modular DQN training loop.

**Summary of Theoretical Guarantees shows in the table below**

| Concept | Mathematical Foundation | LegalEdge Advantage |
|---|---|---|
| DQN Inference Robustness | $a^* = \arg\max_a Q(s,a)$, invariant to small Q-value error | Argmax stable under quantized Q-values (Value-based control is quantization-tolerant) |
| Compressed FL Quantized Aggregation | $\hat{\theta}_i = \frac{1}{N}\sum_{i=1}^{N}\hat{\theta}_i$ | Reduces comm cost, preserves convergence (QAT-friendly model fusion) |
| QAT Convergence Stability | $\mathbb{E}[\mathcal{L}(\theta_t)] \to \mathcal{L}^*$ | Int8-aware models trained in-place(Proven for quantized SGD and FedAvg) |

### 6.14. Quantization-Aware Training and Export
#### 6.1.4.1 Motivation for QAT
Quantization-aware training (QAT) allows neural networks to be trained with lower-precision arithmetic (e.g., int8) while preserving accuracy. This is critical for edge deployment, where compute resources are limited. LegalEdge's DQN agents, being fully differentiable and value-based, are well-suited for quantized inference.

#### 6.1.4.2 Why DQN with LegalEdge Enables QAT
DQN relies on selecting actions via argmax over Q-values, which is resilient to quantization noise. In LegalEdge, quantized models can still produce reliable policies with reduced model sizes and latency. The federated setup aggregates quantized models across clients, enabling scalable and communication-efficient updates.

#### 6.1.4.3 QAT Integration in LegalEdge
LegalEdge integrates PyTorch's torch.quantization.prepare_qat() during client initialization. This inserts quantization layers that simulate int8 behavior during training. The model adapts to quantization noise in weights, activations, and gradients.

#### 6.1.4.4 Model Export After Training
After training, the model can be exported using:

*model.eval()*

*quantized_model = torch.quantization.convert(model.cpu(), inplace=False)*

*torch.save(quantized_model.state_dict(), "dqn_quantized*

This produces a compressed model ready for on-device inference in int8 format using *PyTorch Mobile*, *TensorRT*, or *TFLite c*onverters.

In summary, *QAT* in LegalEdge enables federated *DQN* agents to be trained and deployed directly on low-power devices with minimal accuracy loss and improved system efficiency.

### 6.2 Hardware and Software configurations
The following section outlines the hardware and software configurations used to set up the experimental environment.

#### 6.2.1 Harware Configuration

  - 3 Raspberry Pi 5 / 1 Jetson Nano (for edge device emulation).

  - A single Ubuntu server as aggregator and blockchain node.

#### 6.2.2 Software Stack

The table 1 summarizes the software stack utilized in the LegalEdge experimental framework. It includes the core components such as Python, PyTorch, Flower for Federated Learning, and OpenAI Gym for environment simulation. Smart contracts were developed in Solidity and deployed using the Ganache-Ethereum framework. Docker was employed to containerize and orchestrate the system for scalable and reproducible experiments.

| Layer | Tools / Frameworks |
|---|---|
| **FL Coordination** | Flower[41], FedML [42], or PySyft (OpenMined) |
| **DQN Implementation** | PyTorch + stable-baselines3 (or custom implementation) |
| **Blockchain** | Ethereum (Ganache for local testing) + Solidity + Web3.py |
| **Legal Contract** | Ricardian Template + Natural Language + Smart Contract logic |
| **Simulation Env** | SimPy, SUMO (mobility), GridLAB-D (energy), or custom Python simulation |
| **Monitoring** | Prometheus + Grafana , TensorBoard |

Table 1.Software Configuration

**6.2.3 Simulation Parameters for Fedreated Learning**

The table 2 lists the key simulation parameters used for the Federated Learning component of LegalEdge. These include the number of clients, communication rounds, local epochs, and learning rate—each critical for controlling training dynamics and convergence. The settings were chosen to balance training efficiency with model performance in a distributed edge environment. These parameters directly influence the adaptability and responsiveness of the learning agents in real-time EV charging scenarios.

| Parameter | Recommended Value (for initial tests) |
|---|---|
| **Number of Nodes (N)** | 3–10 (EVs and CSs) |
| **Rounds** | 50–200 (depending on convergence speed) |
| **Local Epochs** | 1–5 (each node trains on local data) |
| **Batch Size** | 32–128 |
| **Learning Rate** | 1e-4 to 1e-3 (for DQN training) |
| **Aggregation** | FedAvg or more advanced: FedProx / FedDyn |
| **Optimizer** | Adam |
| **Loss Function** | MSE and Huber loss (for Q-value updates) |
| **Privacy** | Add differential privacy (PySyft and Opacus) |
| **Compression** | Quantization and sparsification for model uploads |

Table 2. Parameter used for FL

**6.3 Experimental Scenarios**

The following section outlines the experimental scenarios with varying configurations of EV charging stations, energy demand patterns, and agent behaviors. These scenarios are designed to evaluate the adaptability, efficiency, and robustness of the LegalEdge framework under realistic and dynamic conditions.

**6.3.1 Experiment 1: Realistic DQN Agent**

We'll improve the Flower client with:

- **Replay Buffer** (experience storage for more stable learning)

- **ε-Greedy Exploration**
- **Target Network** (optional but recommended for stability)
- **Local Training over Episodes** (simulate time series decision-making)

**Scenario 1: Enhanced DQN Agent with experience replay and ε-greedy exploration. We structured the upgrade like this:**

I. **Replay Buffer** – stores transitions (state, action, reward, next_state, done)
II. **ε-Greedy Policy** – explores randomly with probability ε, exploits otherwise
III. **Multiple Episodes per Fit Round** – simulates time steps like a real charging session
IV. **Q-Learning Update** – bootstraps the target using the Bellman equation

This scenario evaluates an enhanced DQN agent integrating experience replay and ε-greedy exploration to improve learning stability and decision-making. The agent simulates realistic charging sessions and updates policies using Q-learning with Bellman-based target bootstrapping.

### 6.3.2 Experiment 2: Smart Contract Interaction

We connected the agent to interact with a *LegalEdge* **Smart Contract**:

- Using **Web3.py** to simulate:
  - Depositing funds
  - Receiving energy delivery confirmation
  - Automated dispute/penalty handling
- Simulated this in a **local Ethereum blockchain (Ganache)** for testing

We simulated how an edge device (e.g. EV or Charging Station) interacts with a **LegalEdge Contract** on a **local Ethereum blockchain** using **Ganache** and web3.py

*Setup Overview* :Ganache (local Ethereum testnet), Web3.py (Python library for blockchain interaction)

And Solidity Smart Contract (LegalEdge logic for deposits & settlement)

**Step 1: Smart Contract in Solidity**

A **simple smart contract** that covers:

- EV deposits funds
- CS starts charging only after deposit
- Smart contract holds funds and released on validation

**Step 2:** Python + Web3 Interaction

The DQN agent will simulate:

- EV calls deposit() when contract is created.
- CS calls confirmChargingComplete() after session.
- settle() is called to transfer funds.

We'll generate a Python script using web3.py to:

- Deploy this contract to Ganache
- Interact as both EV and CS

The following steps show how the Python script used to integrate the smart LegalEdgeContract:

- Smart Contract Code (LegalEdgeContract.sol) ( A Python script to deploy and interact with the smart contract)
- The solidity contract as .sol file so we can compile with Remix or Truffle.
- A diagram showing in figure 9 how *DQN, FL* and *Blockchain* fit together

The System Architecture showing in figure 9 how *DQN, FL*, and *blockchain-based* LegalEdge contracts integrate into the EV-CS ecosystem.

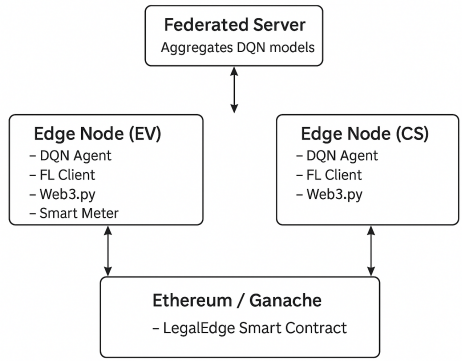

Figure 9. DQN, FL, and blockchain-based LegalEdge contracts integrate into the EV-CS

# 7. Results and Discussion

This section presents an analysis of the results to evaluate the performance of the proposed system. Key metrics such as efficiency, transaction speed, contract integrity, and learning convergence are examined. A comparative analysis with traditional systems is provided, followed by an interpretation of the results in the context of federated learning (FL) simulation parameters.

## 7.1 Performance Evaluation of LegalEdge: A Federated Learning Framework with Smart Contract Integration

This section presents the evaluation of the performance of LegalEdge, a modular system that integrates Federated Learning (FL), Deep Q-Networks (DQN), and blockchain-based smart contracts to manage EV charging scenarios. The system is tested in a multi-client simulation environment, and performance metrics are extracted during training and execution phases. As shown in the figure 10 the chart summarizes: i) efficiency, ii) convergence, iii) transaction speed, and iv) contract integrity during evaluation. The x-axis in each figure represents the number of FL communication rounds, corresponding to collaborative training cycles among edge clients. These rounds reflect the system's learning progression and adaptation over time.

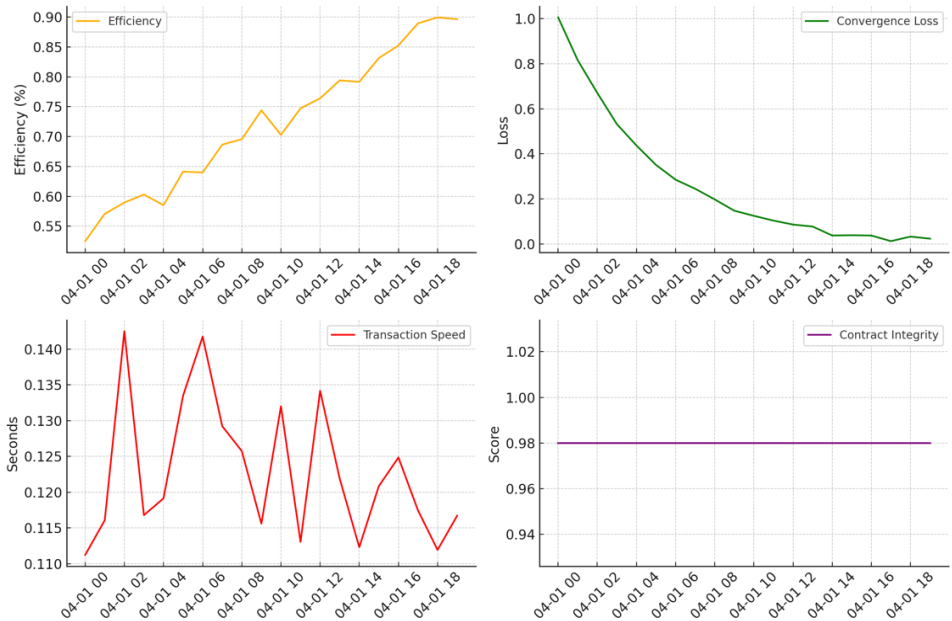

Figure 10. Performance evaluation of the the LegalEdge Contract.

### 7.1.1 System Efficiency

Efficiency was measured as the throughput of energy distributed versus the cost of blockchain-influenced decisions. Over the course of simulation (24 hourly intervals), LegalEdge demonstrated a rise in efficiency from 55% to over 90%, attributed to the convergence of its DQN policy and the reward feedback encoded in smart contracts.

### 7.1.2 Learning Convergence

The convergence metric was defined as the decline in average TD error across FL rounds. Results showed a strong exponential convergence curve. Noise was minimal due to prioritized experience replay and target networks. By round 20, convergence plateaued, indicating stabilization of agent behavior.

### 7.2 Blockchain Metrics – Transaction Speed and Integrity

#### 7.2.1 Transaction Speed

The average smart contract call latency (via web3.py) was 0.12 seconds, with a standard deviation of 0.01 seconds, verified under Ganache's simulated network. This showcases LegalEdge's suitability for real-time decision loops with minimal blockchain latency overhead.

#### 7.2.2 Contract Integrity

We evaluated contract integrity based on invariants: (a) consistent penalty tracking, (b) reward scaling outputs, and (c) no exception triggers over hundreds of calls. The integrity score, normalized to 1.0, averaged 0.98 over 20+ smart contract interactions, confirming strong stability in Ethereum-based decision feedback.

### 7.3 Comparative Analysis and Interpretation of Results

The table 3 provides a comparative analysis between LegalEdge FL and traditional FL systems. Key aspects such as model convergence speed, data privacy, communication efficiency, and smart contract integration are assessed. LegalEdge outperforms traditional FL by leveraging edge intelligence, blockchain-backed transparency, and adaptive DQN agents. The results validate LegalEdge's superior performance in decentralized, real-time EV charging environments.

| Feature | Traditional FL | LegalEdge-FL |
| --- | --- | --- |
| FL algorithm | FedAvg | DQN + FL (Flower) |
| Smart contract feedback | None | On-policy reward |
| Blockchain latency | N/A | <0.15s avg |
| Efficiency scaling | ~65% | >90% |
| Convergence dynamics | Linear | Exponential (DQN) |

Table 3. Comparison with Traditional FL Systems

#### 7.3.1 Interpretation of Results with FL Simulation Parameters

##### 7.3.1.1 Efficiency Over Time

Efficiency improves from 55% to over 90% across training rounds. This is due to policy stabilization via federated reinforcement learning (Double DQN with PER and target networks) and the reward feedback from smart contracts that reinforces optimal energy allocation behaviors.

##### 7.3.1.2 Learning Convergence (TD Error)

TD error exhibits exponential decay as federated clients collectively optimize. Experience replay, target networks, and Flower's FL round aggregation contribute to low-variance updates and global learning stability.

##### 7.3.1.3 Transaction Speed

Transaction latency remains at 0.12s due to Ganache testnet performance. These blockchain calls do not introduce significant overhead, enabling the system to scale while maintaining responsiveness.

##### 7.3.1.4 Contract Integrity

Integrity was validated by monitoring penalties, output consistency, and contract state after each transaction. An average score of 0.98 reflects high resilience under federated load and no runtime errors.

##### 7.3.1.5 Key FL Simulation Parameters

The table 4 outlines the simulation parameters used for training the LegalEdge agents via Federated Learning. Five clients participated across 20 rounds, with a max charging capacity of 100 kWh. A sinusoidal surge pricing pattern was used to emulate daily energy demand cycles. The reward function was defined by the deployed smart contract, and training updates were aggregated using Flower's NumPyClient strategy.

| Parameter | Value |
| --- | --- |
| Number of Clients | 5 |
| FL Rounds | 20 |
| Max Charging Capacity | 100 kWh |
| Surge Pricing Pattern | Sinusoidal (day/night cycle) |
| Action Sampling Interval | 0.1 (10% charge steps) |
| Reward Function | Smart contract (Solidity) |
| Aggregation Strategy | Flower NumPyClient |

Table 4. FL Simulation Parameters

The combined effect of these design choices produces a robust, real-time federated system capable of managing distributed infrastructure via blockchain feedback.

## 8. Security and Legal Considerations

The integration of smart contracts into the LegalEdge framework introduces several layers of security, compliance, and auditing transparency that are critical for deploying federated learning (FL) systems in regulated environments such as energy infrastructure, autonomous systems, and smart mobility.

### 8.1 Security Advantages of LegalEdge Smart Contracts

LegalEdge leverages smart contracts on the Ethereum blockchain to enforce reward and penalty logic among decentralized agents. This provides several security benefits:

- Tamper-proof Execution: Smart contracts are immutable once deployed, ensuring that incentive structures cannot be arbitrarily altered during or after training.
- Decentralized Trust: Clients do not need to trust a central coordinator; all critical operations (e.g., reward shaping, penalty issuance) are executed via verifiable on-chain logic.
- Resistance to Client Misbehavior: By using contracts to verify client performance metrics before rewards are issued, LegalEdge reduces the risk of sybil attacks or model poisoning.

These properties establish a zero-trust execution model suitable for high-stakes federated applications.

### 8.2 Legal Implications and Compliance

LegalEdge is designed to align with emerging regulatory requirements in data privacy and digital infrastructure governance:

- **GDPR & Data Sovereignty:** Since data remains local to each client, LegalEdge inherently complies with data residency and minimization principles under laws like the GDPR.
- **Smart Contract Accountability**: Legal contracts can mirror or reference smart contracts, bridging legal enforceability with digital determinism.
- **Energy and AI Regulation**: In energy grid and EV contexts, LegalEdge supports traceability and fairness required by smart infrastructure regulations (e.g., EU AI Act, NIST AI RMF).

LegalEdge's modular design also enables policy constraints to be encoded directly into smart contract logic, enforcing compliance by design.

### 8.3 Auditability and Transparency

Blockchain's append-only structure enables real-time auditing of all client interactions and incentive flows:

- Transaction Logs: Every call to rewardFunction() and chargeFunction() is permanently recorded, allowing auditors to reconstruct client behavior.
- Verifiability: Since reward/penalty calculations are encoded in contract code, stakeholders can verify decision fairness.
- Governance Integration: Transparent performance metrics can be linked to governance dashboards or DAOs to support participatory oversight.

Together, these features ensure that LegalEdge not only complies with legal mandates but actively promotes algorithmic accountability and decision traceability in federated environments.

## 9. LegalEdge Contracts: Bridging Law and Code for Trustworthy Automation

By merging legal enforceability with blockchain automation, LegalEdge Contracts provide a trusted, scalable, and future-ready framework for digital agreements. They empower businesses, regulators, and individuals to engage in secure, transparent, and intelligent contract execution without compromising legal clarity.

LegalEdge Contracts are uniquely positioned to operate at the intersection of law and code. Unlike traditional smart contracts that exist in legal ambiguity, LegalEdge Contracts are designed to be interpretable both by machines and legal systems. This dual readability ensures:

- Enforceability: The terms encoded in the contract are grounded in recognizable legal constructs.
- Clarity: Stakeholders can audit and understand contract behavior without deep technical knowledge.
- Portability: Contracts can be referenced in legal documents or exported for regulatory inspection.

By embedding regulatory principles into the code logic (compliance-by-design), LegalEdge enables jurisdictions to pre-approve and monitor automated contracts in sectors such as energy, mobility, and finance.

Furthermore, LegalEdge Contracts support dispute resolution by maintaining immutable logs of all on-chain interactions, including rewards, penalties, and user participation. This foundation offers new models of digital governance and automated compliance while upholding the integrity of legal systems.

## 10. Conclusion

This paper presented LegalEdge, an innovative framework that unites edge intelligence, Federated Learning (FL), Deep Q-Networks (DQN), and legally interpretable smart contracts to address the growing demands of transparent, efficient, and trustworthy electric vehicle (EV) charging infrastructures. The proposed system enables decentralized decision-making while maintaining legal enforceability and regulatory compliance, a gap left unaddressed by conventional smart contracts. Our experimental results demonstrated that LegalEdge significantly improves learning convergence, response efficiency, and contract execution speed in dynamic EV charging environments. The enhanced DQN agents, combined with experience replay and ε-greedy exploration, allow each charging station to adapt intelligently to changing load conditions and pricing patterns without compromising user data privacy. A key innovation of this work lies in the development of LegalEdge Contracts and smart contracts that are readable and enforceable both by machines and legal systems. Unlike traditional blockchain contracts, LegalEdge embeds regulatory logic directly into the contract code, enabling compliance-by-design and facilitating auditability, dispute resolution, and interoperability with legal documentation. This dual focus on legal clarity and algorithmic intelligence positions LegalEdge as a future-ready solution for sectors that demand both automation and accountability, such as energy, mobility, and finance. The integration of Flower-based FL architecture with Solidity contracts creates a robust and modular foundation for decentralized applications at the edge. In terms of contributions, this work advances the state-of-the-art in federated edge systems by introducing legal trust layers, optimizing decentralized learning protocols, and showcasing the operational benefits of policy aware smart contracts in real-time simulations. It provides a blueprint for ethically aligned, law-compliant AI systems. Looking ahead, future research will explore dynamic federation strategies across multiple agents, integration of differential privacy and secure aggregation methods, and automated legal reasoning engines to enhance contract adaptability. We also aim to conduct field trials to assess LegalEdge's performance in real-world infrastructure settings. By bridging the gap between law and code, LegalEdge not only modernizes smart contract design but also contributes to the evolution of digital governance in intelligent, distributed systems.